\documentclass[sn-aps,Numbered,referee]{sn-jnl}

\usepackage{amsmath}
\usepackage{graphicx}
\usepackage{dcolumn}
\usepackage{bm}
\usepackage{hyperref}

\usepackage{braket}
\usepackage{ulem}
\usepackage{graphicx}%
\usepackage{multirow}%
\usepackage{amsmath,amssymb,amsfonts}%
\usepackage{amsthm}%
\usepackage{mathrsfs}%
\usepackage[title]{appendix}%
\usepackage{xcolor}%
\usepackage{textcomp}%
\usepackage{manyfoot}%
\usepackage{booktabs}%
\usepackage{algorithm}%
\usepackage{algorithmicx}%
\usepackage{algpseudocode}%
\usepackage{listings}%

\begin{document}

\title[Giant and anisotropic enhancement of spin-charge conversion in double Rashba interface graphene-based quantum system]{Giant and anisotropic enhancement of spin-charge conversion on double Rashba interface graphene-based quantum system}

\author*{\fnm{Alberto} \sur{Anadón$^{*1}$}}\email{alberto.anadon@univ-lorraine.fr}
\author[2]{\fnm{Armando} \sur{Pezo}$^{2}$}
\author[3]{\fnm{Iciar} \sur{Arnay}$^{3}$}
\author[3]{\fnm{Rubén} \sur{Guerrero}$^{3,4}$}
\author[3]{\fnm{Adrián} \sur{Gudín}$^{3,5}$}
\author[1]{\fnm{Jaafar} \sur{Ghanbaja}$^{1}$}
\author[3]{\fnm{Julio} \sur{Camarero}$^{3,5}$}
\author[2]{\fnm{Aurelien} \sur{Manchon}$^{2}$}
\author[1]{\fnm{Sebastien}  \sur{Petit-Watelot$^{1}$}}
\author*[3]{\fnm{Paolo} \sur{Perna$^{*3}$}}\email{paolo.perna@imdea.org}
\author*{\fnm{Juan-Carlos}  \sur{Rojas-Sánchez$^{*1}$}}\email{juan-carlos.rojas-sanchez@univ-lorraine.fr}

\affil[1]{Institut Jean Lamour, Université de Lorraine / CNRS, UMR7198, 54011 Nancy, France}
\affil[2]{Aix-Marseille Université, CNRS, CINaM, Marseille, France}
\affil[3]{IMDEA Nanoscience, C/ Faraday 9, Campus de Cantoblanco, 28049 Madrid, Spain}
\affil[4]{Departamento de física aplicada and INAMOL, Universidad de Castilla la Mancha, 45071, Toledo, Spain}
\affil[5]{Departamento de F\'isica de la Materia Condensada, Instituto Universitario de Ciencia de Materiales "Nicol\'as Cabrera" and Condensed Matter Physics Center (IFIMAC), Universidad Aut\'onoma de Madrid, Campus de Cantoblanco, 28049 Madrid, Spain}











\abstract{

The ever-increasing demand for efficient data storage and processing has fueled the search for novel memory devices. Spintronics offers an alternative fast and efficient solution using spin-to-charge interconversion. In this work, we demonstrate a remarkable thirty-four-fold increase in spin-to-charge current conversion when incorporating a 2D epitaxial graphene monolayer between iron and platinum layers by exploring spin pumping on-chip devices. Furthermore, we find that the spin conversion is also anisotropic. We attribute this enhancement and anisotropy to the asymmetric Rashba contributions driven by an unbalanced spin accumulation at the differently hybridized top and bottom graphene interfaces, as highlighted by ad-hoc first-principles theory. The improvement in spin to charge conversion as well as its anisotropy reveals the importance of interfaces in hybrid 2D-thin film systems opening up new possibilities for engineering spin conversion in 2D materials, leading to potential advances in memory, logic applications or unconventional computing.
}

\keywords{graphene, spin-orbitronics, spin-charge current conversion, spin pumping, Rashba interface}

\maketitle

Interconversion between spin and charge currents is a key operation for modern memories and one of the main research activities in today’s spintronics. 
In general, the generation and detection of spin currents mainly occurs in materials exhibiting large spin-orbit coupling (SOC), such as heavy transition metals like Pt \cite{Miron2011,Liu2011a} or Ta  \cite{Liu2012}.
Lately, it has been realized that two-dimensional (2D) systems with inversion symmetry breaking, such as Rashba interfaces \cite{Manchon2015,Rojas-sanchez2013_Ag-Bi,Lesne2016} and the surface of topological insulators \cite{mellnik2014spin, Rojas-sanchezPRL2016}, display much larger spin-charge interconversion efficiencies compared to transition metals \cite{Rojas-Sanchez2019ComparedSystems}. This superiority stems from the strong spin-momentum locking of the 2D states such as Rashba surface states induced by the strong interfacial potential gradient. Among the promising 2D systems for spin-charge interconversion \cite{galceran2021control}, graphene (Gr) stands out as a particularly appealing platform. 

As a matter of fact, although pristine Gr is a poor spin-charge converter due to its vanishingly small intrinsic SOC, its remarkable ability to acquire a sizable SOC by proximity effect and doping has been widely reported \cite{SaveroTorres2017}. It has been demonstrated for example that the intercalation of either (non-magnetic or magnetic) transition or rare-earth metals underneath a Gr monolayer grown on single crystal substrates results in large magnetic anisotropy \cite{Shikin2013, Blanco-Rey2021, Rougemaille2012,Ajejas2018}, spin-polarization, Rashba states \cite{marchenko2012giant} and Dzyaloshinskii–Moriya interaction \cite{Yang2018, Ajejas2018,cano2023rashbalike}, or even in the appearance of non-trivial, exotic topological spin textures and emerging symmetry-broken phases \cite{Jugovac2023}. In the context of spin-charge interconversion, taking advantage of the exceptional conductivity of Gr monolayers and their ability to acquire SOC by proximity to other materials opens avenues for the design of efficient spin-charge interconverters. 

The physics of spin-charge interconversion in 2D systems is markedly different from that of metallic heterostructures. In the latter, the interconversion is governed by the inverse spin Hall effect (ISHE) \cite{Sinova2015}, attributed to spin-dependent extrinsic scattering or intrinsic anomalous velocity, that converts a spin {\it current} into a charge current. In the former, the interconversion is induced by the so-called inverse Edelstein effect (IEE), associated with the helical spin texture and that converts a spin {\it density} into a charge current \cite{Edelstein1990, Rojas-sanchez2013_Ag-Bi, Lesne2016}. 
The spin-charge interconversion can be probed by spin-orbit torque (SOT) (charge-to-spin) or by spin-pumping (spin-to-charge) techniques, the latter being unaffected by shunting in other metallic layers for the quantification of the reciprocal effects, ISHE and IEE \cite{Rojas-Sanchez2019ComparedSystems}. The figure of merit for the IEE efficiency has the unit of length, and its effective value will be denoted by $\lambda$. \cite{Rojas-sanchez2013_Ag-Bi, Lesne2016, Rojas-sanchezPRL2016} 

In exfoliated  Gr on diverse supports, many spin transport experiments devoted to reveal spin relaxation, diffusion, precession \cite{Tombros2007_Nature,Tombros2008}, and induced SOC by proximity with heavy metal \cite{SaveroTorres2017} or 2D materials \cite{Safeer2019,Ghiasi2019,Kamalakar2014_Gr,Anderson2023} have been reported. However, in all cases, ordinary Hall effect and shunting in the metallic layers \cite{Safeer2021} limit the straightforward and reliable quantification of the effects. 

In pristine Gr the spin relaxation time may change between the in-plane, $\tau_s^\parallel$, and out-of-plane, $\tau_s^\perp$, spin directions \cite{Tombros2008}. Such anisotropy can be enhanced by proximity with heavier materials, typically reaching $\tau_s^\perp\sim 30 {\rm ps}\gg\tau_s^\parallel\sim3$ ps for Gr/WS$_2$ \cite{Benitez2018StronglyTemperature}. 
Instead, no spin relaxation anisotropy along different {\it in-plane} propagation directions has been reported.

Regarding spin-charge interconversion, by using spin pumping measurements very different efficiencies were measured by interfacing Gr with other materials. While $\lambda\approx0.002$ nm   was observed in pristine Gr \cite{Mendes2019}, Gr on a ferrimagnetic insulator garnet YIG showed  $\lambda\approx0.001$ nm, and larger efficiencies were measured in Gr on Ni$_{81}$Fe$_{19}$ $\approx0.003$ nm and on YIG/MoS$_2$ \cite{Mendes2019}. 
However, most of these experiments were performed either in exfoliated flakes of Gr and other van der Waals 2D materials \cite{Yan2017} or by using Cu substrate and PMMA resist. In all these approaches, it is extremely difficult to ensure atomic control on the Gr structure and its interfaces, and avoiding possible contamination is challenging. This leads to devices with unrealistic results to show the real potential of Gr in contact with other materials. Particularly to show the ability of graphene to acquire SOC or to form an efficient magnetic Rashba interface for interconversion when in contact with a magnetic layer.

In this work, we resort to our ability to prepare epitaxial heterostructures in which a Gr single layer, ferromagnetic and heavy metallic ultra-thin layers are stacked. Such Gr-based heterostructures, deposited on insulating oxide supports to enable transport experiments \cite{Anadon2021}, are fabricated under ultra-high-vacuum (UHV) conditions and in-situ analyzed to guarantee structural atomic perfection and contaminant-free surfaces \cite{Ajejas2018,Ajejas2020,Blanco-Rey2021}. 
Thus, taking advantage of the unique ability of Gr to acquire proximate SOC and mediate spin currents, we demonstrate a massive enhancement of the spin-charge interconversion in a transition metal bilayer due to the insertion of atomically thick epitaxial Gr. 
The quantification of the conversion of charge currents into spin currents in different systems, including control samples, allows us to unambiguously assess the overall efficiency of the reciprocal effect, spin-to-charge current conversion. 
We find that the Fe/Gr/Pt system has a charge current production with a gain of at least 17-fold compared to Fe/Pt. This system exhibits a large anisotropy of spin current to charge current conversion. The gain of the charge current production in the more efficient direction increases to about 34 times that of Fe/Pt. This result contrasts sharply with the case of Co/Gr/Pt, where a reduction of the overall effective efficiency was previously observed using the spin Seebeck effect \cite{Anadon2021}. Our theoretical results based on first principles calculations show that Co/Gr/Pt has an effective interfacial spin-orbit coupling opposite to that of Fe/Gr/Pt. This is equivalent to an effective spin Hall angle $\theta_{SHE}$ (or effective IEE length $\lambda $) of opposite signs at Co/Gr and Fe/Gr interfaces. Accordingly, our theoretical results reconcile our surprising experimental results on Fe/Gr/Pt and Co/Gr/Pt. Remarkably, we report an overall IEE efficiency of about 4.8 nm for the more effective direction, $\Gamma-M$, of Fe/Gr/Pt, which is among the highest values reported so far at room temperature. Concurrently, we report a strong in-plane spin-charge current conversion anisotropy with an IEE efficient of about 0.27 nm for the $\Gamma-K$ direction.

\begin{figure}[!h]%
\centering
\includegraphics[width=\textwidth]{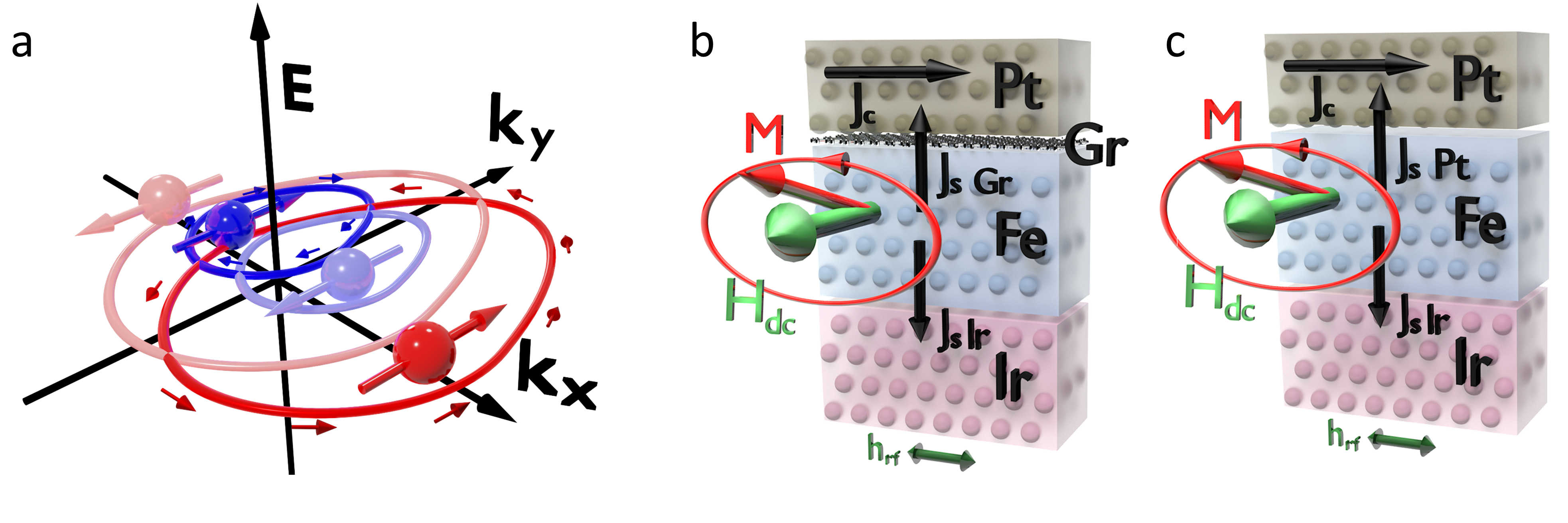}
\caption{ \textbf{Ir/Fe/Gr/Pt stack for spin pumping and Inverse Edelstein Effect}. \textbf{(a)} Schematic of a magnetic Rashba interface in the reciprocal space at the Fermi level under a DC-applied magnetic field along $y$, and their displacement due to a spin current injection. There are two Fermi contours with opposite spin chiralities, clockwise and contour-clockwise. If a spin current, with spin polarization along $y$, is injected, it shifts the outer Fermi contour along $k_x$ and the inner Fermi contour in the opposite directions generating a net transverse charge current. \textbf{(b,c)} 3D view of the stacking of our main system //Ir/Fe/Gr/Pt \textbf{(b)} containing the double Rashba interface Fe/Gr/Pt; and the Gr-free control system \textbf{c}.  
}\label{fig:Intro}
\end{figure}

\textit{\bf Inverse Edelstein effect at magnetic Rashba interface}. 
Fig. \ref{fig:Intro} schematically presents the unbalanced spin accumulation due to asymmetric Rashba contributions occurring at Fe/Gr and Pt/Gr interfaces . 
The two Rashba contours in the reciprocal space of a magnetic Rashba interface shown in Fig. \ref{fig:Intro} at the Fermi level are deformed by the application of a DC magnetic field along the $y$-direction. 
The outer contour has a counterclockwise spin polarization helicity opposite to that of the inner contour. When a spin current with a spin polarization along $y$ is injected along the $z$ direction, $J^z_{s^y}$, the spin population is increased by $\delta s^y$. This could happen only around $+k_x$ due to the counterclockwise helical spin texture of the outer Fermi contour. On the contrary, the spin population is reduced by $-\delta s^y$ around $-k_x$ due to the spin helicity of the external contour. Consequently, the external contour is shifted $+\delta k_x$ towards $+k_x$. 
Since the inner contour has the opposite helical spin texture, it is shifted by $-\delta k_x'$ towards $-k_x$. 
This results in a net effect of spin accumulation $\delta s^y$ along $k_x$ direction as displayed in Fig. \ref{fig:Intro}a. A variation of electron momentum $\delta k_x$ means that an electric or charge current is being generated laterally and transverse to the direction of injected $J^z_{s^y}$. 
The generated charge current, in this quantum system with double interface Fe/Gr and Pt/Gr as schematized in Fig. \ref{fig:Intro}a, can be experimentally detected by measuring the DC spin pumping voltage. A schematic of the main system at the magnetic resonance condition is illustrated in Fig. \ref{fig:Intro}b, and of the Gr-free control system in Fig. \ref{fig:Intro}c.In an open circuit, we measure the voltage due to the production of the charge current.

\section*{Results}

\subsection*{Growth and structural characterization of Gr-based double Rashba interface}\label{subsecGrowth}
In order to realize high-quality epitaxial heterostructures in which a Gr monolayer is embedded in between a top heavy metal and a bottom ferromagnetic layers, we have resorted to the procedure detailed in the Methods, which has been optimized from the case of Gr/Co \cite{Ajejas2018, Ajejas2020, Blanco-Rey2021} . 
In brief, for samples with graphene we first prepared a 10 nm-thick epitaxial Ir(111) buffer grown by DC sputtering on commercially available Al$_2$O$_3$(0001) single crystal substrates at 670 K, on which an epitaxial Gr monolayer was grown by ethylene chemical vapor deposition (CVD) at 1025 K under a partial pressure of $5.5\cdot10^{-6}$ mbar during 30 min. 
The Fe film was then evaporated on top of graphene by molecular beam epitaxy (MBE) at room temperature (RT) in UHV and at low deposition rate. The temperature was gradually raised to less than 550 K in order to activate the Fe intercalation process while avoiding unwanted metal intermixing. Between 12 and 14 nm-thick homogeneous Fe layers with high structural order and well-defined interfaces were thus obtained (see Fig. S2 in Suppl Mat.). 
Concurrently, reference Ir/Fe samples without graphene were prepared depositing Fe by DC sputtering at RT on top of Ir. Finally, in all samples, a 5 nm-capping layer of Pt or Al was DC sputtered at RT. In the following we will refer to the Ir/Fe/Gr/Pt stack as the main system, Ir/Fe/Pt as a reference, while the Ir/Fe/Al and Ir/Fe/Gr/Al stacks are the control samples.

\begin{figure}[!h]%
\centering
\includegraphics[trim={0 0 0 0},width=\textwidth]{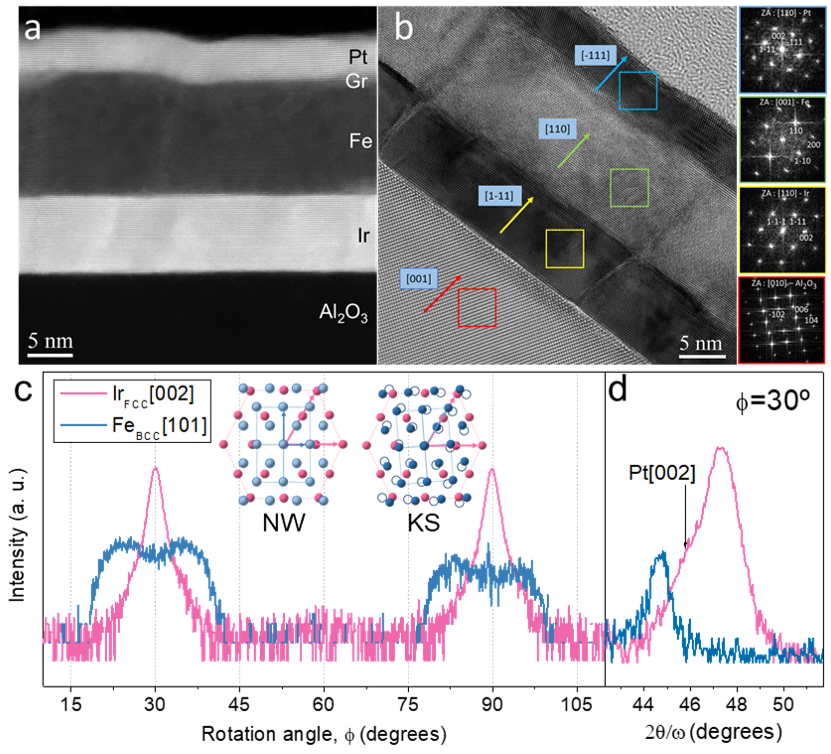}
\caption{ \textbf{Structural characterization}. {\bf (a)} High-angle anular dark field scanning transmission electron micrograph cross section images of the Ir/Fe/gr/Pt sample. Close up of graphene interface probing flat and well defined interfaces. {\bf (b)} High resolution transmission electron micrograph and corresponding Fast Fourier Transform (FFT) in the Ir, Fe and Pt regions respectively. FFT patterns probed [110] oriented growth of the Fe layer with BCC structure. Pt on top of graphene shows [111]-oriented epitaxial growth. {\bf (c)} $\phi$ scans around Ir and Fe reflections show a 6-fold symmetry, indicating the coexistence of equivalent in plane twin boundary domains for Ir(111) as well as Nishiyama-Wassermann (NW) and Kurdjumov-Sachs (KS) configurations for Fe$^{\rm BCC}$ coupling with the Ir$^{\rm FCC}$ structure. {\bf (d)} 2$\theta$ scans at [002]$_{\rm Ir}$($2\theta=47.29^{\circ}$) and [101]$_{\rm Fe}$($2\theta=44.67^{\circ}$) reflections indicate inconmensurate growth of both layers, with bulk like lattice parameter. Asymmetric profile for Ir reflection is related with the presence of pseudomorphic Pt on top of Gr.}\label{fig:structure}
\end{figure}

The results of the structural characterization are shown in Fig. \ref{fig:structure}. 
High resolution Scanning Transmission Electron Microscopy (STEM) and X-ray Diffraction (XRD) characterization prove a successful growth of the epitaxial Ir and Fe layers in both reference and control samples. 
In particular, sharp interfaces between the layers are clearly observed in STEM micrographs in Fig. \ref{fig:structure}a, whereas the high crystallographic order together with atomic resolution is shown in Fig. \ref{fig:structure}b as well as in the XRD 2$\theta/\theta$ in Fig. S1 of Suppl. Mat. 
In the latter we see that the Ir atoms grow along the [111] crystallographic direction with a face-centered structure (FCC) on top of Al$_2$O$_3$[0001] substrates. 
The low roughness and homogeneity of the interfaces is also confirmed by the X-Ray Reflectivity (XRR) measurements (see Fig. S1 in Supp. Mat.).

From a careful check of the XRD, we observe that instead of a typical 3-fold in-plane symmetry, $\phi$-scans around [002]$_{\rm Ir}$ crystallographic reflection shown in Fig. \ref{fig:structure}c presents rather a 60$^{\circ}$ periodicity. This is due to the presence of two in-plane configurations rotated by 180$^{\circ}$. Thus, the following crystallographic relations are found: [0$\bar{1}$10]$_{\rm{Al_2O_3}}\|$[1$\bar{1}$0]$_{\rm Ir}$ and [0$\bar{1}$10]$_{\rm{Al_2O_3}}\|$[$\bar{1}$10]$_{\rm{Ir}}$ \cite{Ajejas2020, Anadon2021, Blanco-Rey2021}. 
It is known that CVD Gr grows axis on axis on Ir, i.e., [0$\bar{1}$10]$_{\rm Al_2O_3}\|$[1$\bar{1}$0]$_{\rm Ir}$$\|$[11$\bar{2}$0]$_{\rm Gr}$ \cite{Blanco-Rey2021}, 
and that presents a characteristic $10\times10$ moir\'e pattern due to incommensurate lattice of Ir and Gr \cite{Ajejas2020}. 

Upon intercalation, Fe grows incommensurate along the [110] with a base-centered cubic (BCC) structure on top of FCC Ir stacked along [111], as commonly observed for the Fe/Ir system with $t_{\rm Fe}>$ 3 ML  \cite{10.1063/1.4919123,hsieh2021strain}. BCC Fe stacked along [110] presents a tetragonal structure. In the $\phi$-scans acquired around the [101]$_{\rm Fe}$ shown in Fig.\ref{fig:structure}c we can observe a 60$^{\circ}$ periodicity that is due to the coexistence of different in-plane configurations for the coupling between FCC Ir and BCC Fe structures indicated in the insets: [$\bar{1}$01]$_{\rm Ir}$$\|$[001]$_{\rm Fe^{\rm BCC}}$ (Nishiyama-Wasserman); [$\bar{1}$01]$_{\rm Ir}$$\|$[1$\bar{1}$1]$_{\rm Fe^{\rm BCC}}$ (left Kurdjumov-Sachs) and [$\bar{1}$01]$_{\rm Ir}$$\|$[$\bar{1}$11]$_{\rm Fe^{\rm BCC}}$ (right Kurdjumov-Sachs) \cite{ohtake2008epitaxial}.
Fig. \ref{fig:structure}d shows the 2$\theta$ scans at [002]$_{\rm Ir}$ and [101]$_{\rm Fe}$ reflections. Maximum intensity is found at $2\theta$=47.29(1)$^{\circ}$ and $2\theta$=44.67(1)$^{\circ}$ for Ir and Fe respectively, which is consistent with a fully relaxed growth of both structures. [002]$_{\rm Ir}$ reflection show an asymmetric profile related with the presence of pseudomorphic Pt on top of Gr.

\subsection*{Giant and anisotropic spin-charge current interconversion}\label{sec:SCC}
\begin{figure}[!h]%
\centering
\includegraphics[width=\textwidth]{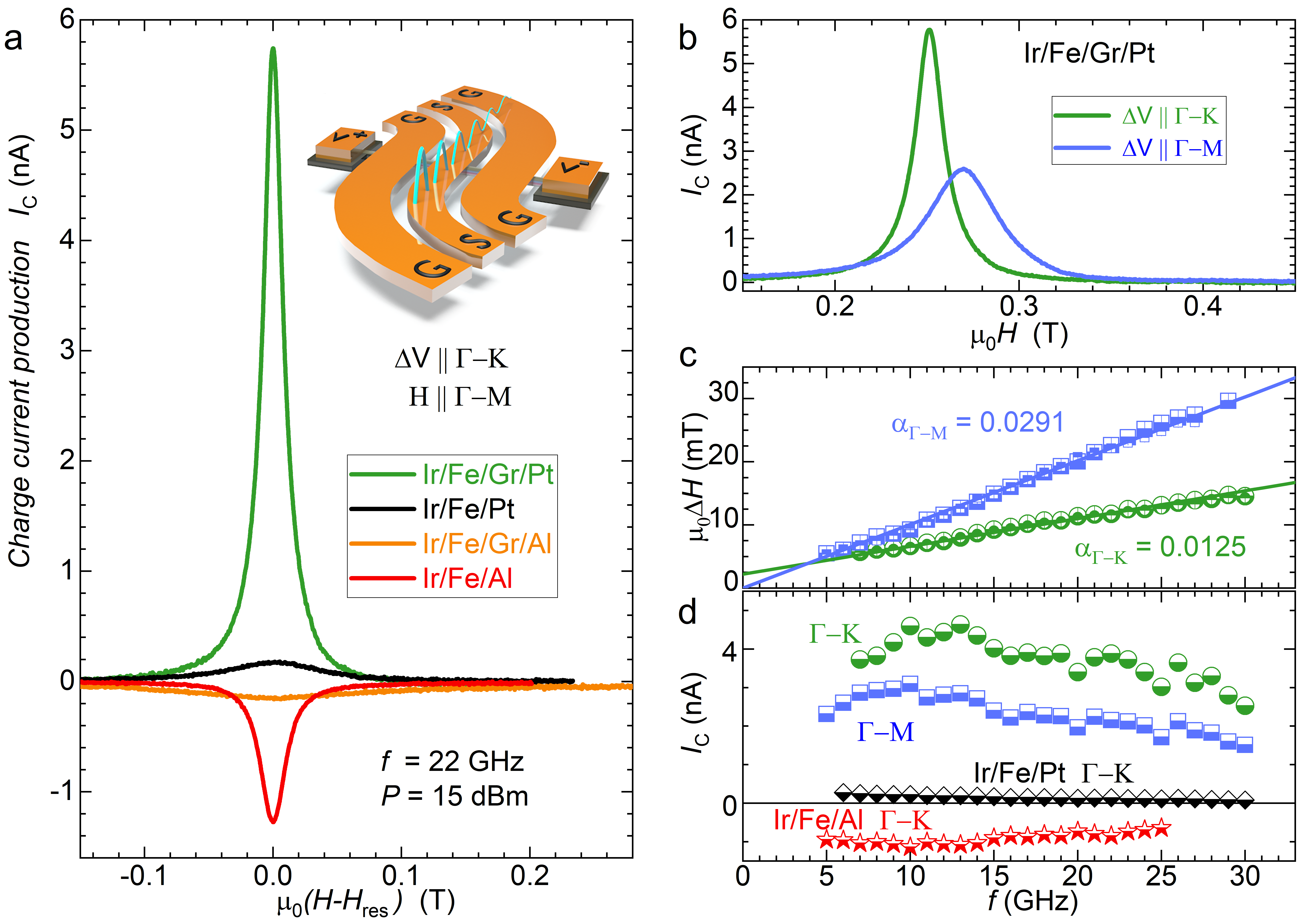}
\caption{\textbf{Results of spin-to-charge current conversion}.  \textbf{(a)} Raw data of $V_{sp}$ normalized by the total resistance of the measured bar shown for 4 systems and when $H||$$\Gamma-M$ and $\Delta V||$$\Gamma-K$. The iridium contribution predominates in the Ir/Fe/Al trilayer. Then the signal is almost totally canceled in Ir/Fe/Gr/Al. The Pt signal predominates over Ir in Ir/Fe/Pt. The signal is strongly enhanced when Gr is sandwiched between Fe and Pt, Ir/Fe/Gr/Pt, with a 34-fold gain over Ir/Fe/Pt. In those cases, the spin pumping bar was etched along the $\Gamma-K$ or $[11\bar{2}0]$ direction of the sapphire substrate. Charge current production $I_c$ \textbf{(b)}, and bandwidth frequency dependence of the linewidth \textbf{(c)} along two non-equivalent in-plane directions of high symmetry for Ir/Fe/Gr/Pt. 
We can observe an anisotropy in amplitude and linewidth \textbf{(b)}, \textit{i.e.}, a clear magnetic damping anisotropy \textbf{(c)}. \textbf{(d)} $f$-dependence of the charge current produced according to $\Delta V$ directions in the different systems studied. Damping and charge current production are clearly different for the different directions only for Ir/Fe/Gr/Pt. This suggests an anisotropy of interconversion due to the Rashba states at Fe/Gr/Pt interfaces. The other systems have isotropic damping and charge current production. See also Suppl. Information.}\label{fig:SP}
\end{figure}

As mentioned above, our chosen method to study and quantify the reciprocal effects, ISHE and IEE, is based on spin-pumping ferromagnetic resonance (SP-FMR). SP-FMR measurements are performed in micrometric devices composed of the sample slab with a width of $W= 10 \mu$m, an insulating 200 nm-thick SiO$_2$ layer separating the sample, and a coplanar waveguide (CPW) patterned on top \cite{Fache2020,Arango2022b} as described in the Methods. 
At the resonance condition, Fe magnetization precesses and an out-of-equilibrium spin accumulation occurs at the interfaces with the Fe layer which in turn can diffuse as a spin current in adjacent systems \cite{Tserkovnyak2002a}. 
The injected spin current is converted into charge current by the ISHE in the top Pt or bottom Ir layers, or by the IEE in Rashba states, if any, at the interfaces. 
It has been shown that Ir has the same ISHE sign as Pt, with a much lower efficiency (of about 25\% of that of Pt) \cite{Fache2020}. Thus, the total charge current of our main system will be given by the contributions of the ISHE of Pt \cite{rojas2014spin,Fache2020} and Ir \cite{Fache2020}, the latter being of opposite sign since it is located below Fe (Fig. \ref{fig:SP}a and b), and the contribution of the IEE \cite{Rojas-sanchez2013_Ag-Bi,Rojas-sanchezPRL2016} due to the Rashba states, Fig. \ref{fig:SP}b.
We divide the SP-FMR voltage amplitude (\textit{V}\textsubscript{SP-FMR}) by the total resistance of the slab $R$ \cite{Fache2020,rojas2014spin} to get the total charge current produced (\textit{I}\textsubscript{c}). We then have

\begin{equation}
\label{eq:SPIc}
\begin{split}
    I\textsubscript{C}/W & = V\textsubscript{SP-FMR}/RW \\
    & = J_{s,Gr}\lambda+J_{s,Pt}\theta_{Pt}l_{sf}^{Pt}\tanh(t_{Pt}/2l_{sf}^{Pt})-J_{s,Ir}\theta_{Ir}l_{sf}^{Ir}\tanh(t_{Ir}/2l_{sf}^{Ir}),
    \end{split}
\end{equation}
where $l_{sf}^{Pt,Ir}$ is the spin diffusion length, $\theta_{Pt,Ir}$ the effective spin Hall angle, and $t_{Pt,Ir}$ is the thickness of Pt or Ir, respectively. The effective spin current density injected at the Gr interface, $J_{s,Gr}$, is given by \cite{Fache2020,rojas2014spin}

\begin{equation}
\label{eq:Js}
J_{s,Gr}=\frac{2e}{\hbar}\frac{g_{\uparrow\downarrow}\gamma^2 (\mu_0 h_{rf})^2 \hbar}{8\pi\alpha^2}\frac{\gamma\mu_0 M_{eff}+\sqrt{(\gamma\mu_0M_{eff})^2+4\omega^2}}{(\gamma\mu_0 M_{eff})^2+4\omega^2},
\end{equation}
where $e$ is the electron charge, $\gamma$ the gyromagnetic ratio, and $\omega=2\pi f$ is the microwave pulsation. $g_{\uparrow\downarrow}$ stands for the real part of the effective spin mixing conductance. 


\textbf{\textit{Giant enhancement}}. 

A summary of our main experimental results of spin-to-charge conversion is given in Fig. \ref{fig:SP}a where the charge current production $I_{c}$ is reported as a function of $(H-H_{res})$ for four different systems at a spin pumping frequency of $f=22$ GHz. 
Here, the DC magnetic field $H$ is set along $\Gamma-M$ direction of sapphire substrate while the DC voltage is collected in the orthogonal direction, $\Gamma-K$. 
By looking at Fig. \ref{fig:SP}c, we first observe that Ir/Fe/Al sample has a negative signal (red line), -1.28 nA, which is consistent with $\theta_{Ir}>0$ \cite{Fache2020}. Since the contribution of Al is negligible as it has a weak SOC and serves only as a capping layer, the charge current is exclusively generated at the Ir/Fe interface. 
Once Gr is introduced between Fe and Al, i.e., in Ir/Fe/Gr/Al, the signal practically disappears, -0.15 nA (orange line). 
We attribute this to the Fe/Gr magnetic Rashba interface that almost completely balances the ISHE effect of Ir. 
We can thus conclude that $\lambda_{{Fe/Gr}}$ has the same sign as Ir and Pt, and its value is slightly lower than the overall efficiency of Ir, $\lambda_{{Fe/Gr}}\leq \theta_{Ir}l_{sf}^{Ir} \approx 0.03 $ nm \cite{Rojas-Sanchez2019ComparedSystems, Fache2020}. 

Replacing now Al by Pt, the produced charge current is positive in the reference sample Ir/Fe/Pt (black line), 0.17 nA. This is also consistent with the fact that Ir has the same sign of $\theta_{Pt}$ but is less efficient than Pt \cite{Fache2020}, so $\theta_{Ir}l_{sf}^{Ir}<\theta_{Pt}l_{sf}^{Pt}\approx 0.2 $ nm \cite{rojas2014spin,Rojas-Sanchez2019ComparedSystems}. 
However, the signal is greatly increased to a value of 5.74 nA (green line) when Gr is inserted, i.e., in the double Rashba interface quantum system Ir/Fe/Gr/Pt. 
This is 34 times the charge production in Ir/Fe/Pt under the same spin pumping conditions. 

This giant enhancement is consistent with the scenario that Fe/Gr contributes with a positive $\lambda$ as deduced by comparing the signals from Ir/Fe/Gr/Al and Ir/Fe/Al samples. Furthermore, the Pt layer deposited on top of Gr imprints a large SOC at the Gr interface which was not the case with Al's protective coating. Now, both Fe/Gr and Gr/Pt interfaces contribute with the same sign to the spin-to-charge conversion. Consequently, the Fe/Gr/Pt double interface is largely more efficient than Pt: $\lambda_{{Fe/Gr/Pt}}\gg\theta_{Pt}l_{sf}^{Pt} \approx 0.2 $ nm. 

Quantifying the individual contributions of each layer and interface can be a complicated puzzle. Before tackling the quantification, we will discuss the anisotropic behaviour of Fe/Gr/Pt. 

\textbf{\textit{Anisotropic damping, $g_{\uparrow\downarrow}$, and conversion in Fe/Gr/Pt}}. 
Beyond the experimental error bars, Fig. \ref{fig:SP}b shows that the charge production along the $H||\Gamma-M$, $\Delta V||\Gamma-K$ direction, 2.59 nA, is lower than the one along the $H||\Gamma-K$, $\Delta V||\Gamma-M$ direction, 5.74 nA. This anisotropy has been confirmed for all measured frequencies in both the damping, Fig. \ref{fig:SP}c, and the charge current production, Fig. \ref{fig:SP}d.
The observed anisotropy does not occur within the Gr-free system such as Ir/Fe/Pt and Ir/Fe/Al where, within the error bars, the studied directions present similar results. 
By performing measurements on slabs patterned every 30$^\circ$  (see Extended Figure A in Methods, and Supplemental information), we observe a clear anisotropic I$_{\rm C}$ signal in samples with Gr/Pt. This indicates that Gr imprints on Fe a 6-fold anisotropy, which is corroborated by the theoretical calculations shown in the next section.

To properly quantify the spin-charge conversion efficiency and determine $\lambda_{{Fe/Gr/Pt}}$, given the strong efficiency of the double interface, we can consider that all the conversion above the Fe layer happens at the Fe/Gr/Pt double Rashba interface, and no spin current reaches the Pt layer. 
This is also confirmed by our theoretical results.
The strength of the microwave field on the sample at 22 GHz was experimentally determined to be $\mu_0 h_{rf}=0.055$ mT, (see Suppl. Inf. for some details and reported elsewhere), and considering $g_{\uparrow\downarrow}^{\Gamma-K}\simeq 3\times10^{18}$  m$^{-2}$, we have $J_{s Ir/Fe/Gr/Pt}=0.15$ MA/m$^2$. 
The total current produced in our Fe/Gr/Pt double interfaces is 7.02 nA, so it yields an overall IEE efficiency $\lambda_{{Fe/Gr/Pt}}^{\Gamma-K}=4.8$ nm for the more efficient direction (average of measurements between 5 GHz and 23 GHz). In the other direction, considering that $g_{\uparrow\downarrow}^{\Gamma-M}\simeq 2\times10^{20}$  m$^{-2}$, we have an overall IEE efficiency $\lambda_{{Fe/Gr/Pt}}^{\Gamma-M}=0.27$ nm, which is similar to the one of Pt, $\theta_{Pt}l_{sf}^{Pt}\approx 0.2 $ nm. 

Note that it is not possible to simply use a reference damping $\alpha_0$ to obtain $g_{\uparrow\downarrow}$ due to the anisotropy. Instead, a dependence on the Fe-thickness of the main system, Ir/Fe/Gr/Pt, should be performed. This is beyond the scope of the present study. However, the effective values of $g_{\uparrow\downarrow}$ used are conservative. For example, we see that even in the less efficient direction, $I_C=2.59$ nA, which is clearly more efficient than Ir/Fe/Pt, 0.15 nA. This giant efficiency is at room temperature, and it is comparable with the result for the $\alpha$-Sn topological insulator, $\lambda_{{\alpha-Sn}}=2$ nm \cite{Rojas-sanchezPRL2016}. 
Furthermore, the great advantage of our new system is that it is robust for the fabrication of micro and nanodevices. Our spin pumping measurements were performed entirely on-chip, which remains difficult to achieve in $\alpha$-Sn deposited on a fragile substrate like InSb or a little toxic like CdTe.

The anisotropic effective efficiency $\lambda$ reported here might be due Rashba coupling parameter, $\alpha_R$, or the relaxation times $\tau$, both linked to the SOC of the system, $\lambda=\alpha_R\tau/\hbar$ \cite{Rojas-sanchez2013_Ag-Bi}. So, along the two directions studied we have that $\lambda^{\Gamma-K}/\lambda^{\Gamma-M}\approx18$. Theoretical modeling might shed light if $\alpha_R$ or $\tau$ dominates the experimentally measured anisotropy.

\subsection*{First principles analysis of the spin-charge interconversion}\label{sec:theory_cal}

\begin{figure}[!h]%
\centering
\includegraphics[width=\textwidth]{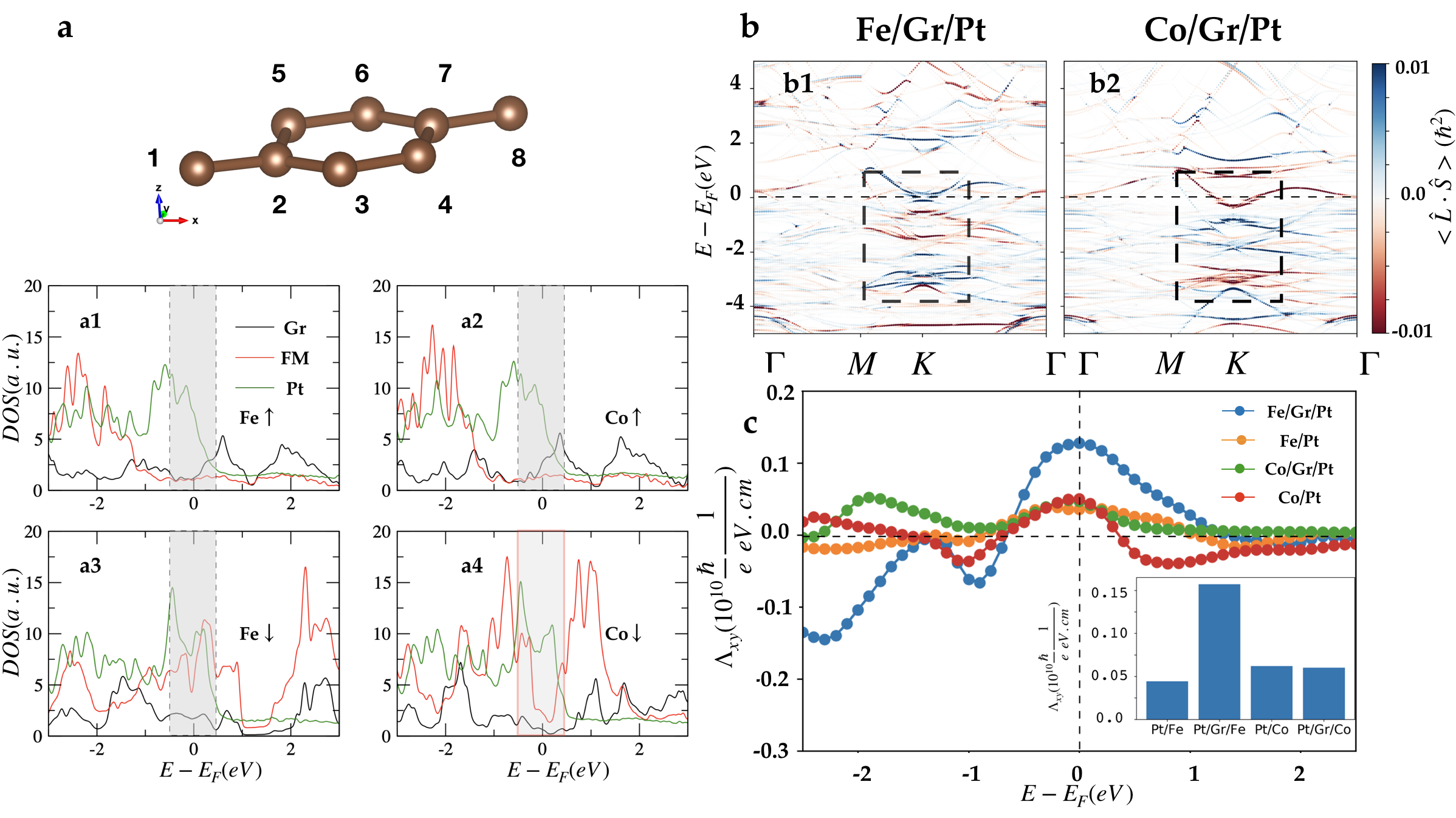}
\caption{ \textbf{Theoretical calculation}. \textbf{(a)} Sketch of the Gr layer showing the carbon atoms' positions. First-principles calculation of projected DOS for minority (a1, a2) and majority (a3, a4) spin populations. The left (a1, a3) and right panels (a2, a4) correspond to Fe/Gr/Pt and Co/Gr/Pt, respectively. Although the spin quantum number is strictly non-conserved in this case, we consider the diagonal parts of the full DOS which are two orders of magnitude larger than the off-diagonal contributions. \textbf{(b)} Spin-orbit coupling projected on Gr for Fe/Gr/Pt (b1) and Co/Gr/Pt (b2). The enhanced SOC of Gr, induced by the proximate transition metals, changes sign when changing Fe for Co, which indicates a change in the sign of the spin-to-charge conversion in these heterostructures. This is confirmed by calculating explicitly the Rashba-Edelstein tensor projected on the Pt layer next to the interface. \textbf{(c)} Electrically-driven effect calculated for heterostructures (Fe,Co)/Gr/Pt for blue and green curves, (Fe,Co)/Pt in orange and red curves. The inset shows the values for the Rashba-Edelstein tensor calculated at the Fermi level. 
}\label{fig:Theory}
\end{figure}

To explain our experimental results, we have performed first principles calculations of two trilayers, Fe(2 ML)/Gr/Pt(7 ML) and Co(2 ML)/Gr/Pt(7 ML), following the procedure described in the Methods. The density of states (DOS) projected on each layer is reported in Figs. \ref{fig:Theory}(a1-a4). We show the contributions of the majority (a1, a3) and minority spin populations (a2, a4) for Fe/Gr/Pt (a1, a3) and Co/Gr/Pt (a2, a4). Whereas the projected DOS for the majority spins is similar for the two systems, the DOS of the minority spins of Fe at the Fermi level (a3) is larger than that of Co (a4). Concomitantly, the minority spins of Gr have a reduced DOS in proximity with Co compared to Fe. This difference is reflected in the magnitude of the spin moment induced on the Gr monolayer, which amounts to -1 $\mu_B$ in Fe/Gr/Pt and -0.6 $\mu_B$ in Co/Gr/Pt (see Table \ref{table:mag_moment_graphene} the Methods), manifesting a substantially different hybridization scheme. Most remarkably, the SOC induced on Gr by proximity to the transition metals is strikingly different in the two systems, as displayed in Fig. \ref{fig:Theory}b for Fe/Gr/Pt (b1) and Co/Gr/Pt (b2). We find a clear inversion of the sign of the induced SOC: near the Fermi level, $\braket{\mathbf{l}\cdot \mathbf{s}}> 0$ in Fe/Gr/Pt, whereas $\braket{\mathbf{l}\cdot \mathbf{s}}< 0$ in Co/Gr/Pt. 

To assess how this sign reversal influences the overall spin-charge interconversion process, we compute the direct Edelstein effect, i.e., the electrical generation of non-equilibrium spin density. Based on Onsager reciprocity, the IEE is directly proportional to the direct Edelstein effect \cite{manchon2024spin}. The current-driven spin density is computed within the linear response formalism considering the symmetrized decomposition of the Kubo-Bastin formula \cite{varga_2020}. In the present situation, we applied an electric field along $\hat{x}$ and computed the components of the spin operators along $\hat{y}$.
The results are reported in S13 (Supplementary information) where is depicted the layer projection of the spin accumulation calculated at Fermi level, showing at the same time the sign reversal of the direct Edelstein effect in Gr, consistent with the sign reversal of the SOC reported in Fig. \ref{fig:Theory}b. In order to quantify the direct Edelstein effect we have calculated the Rashba-Edelstein tensor as a function of the chemical potential shown in Fig. \ref{fig:Theory}c. We note that the presence of Gr enhances the response, particularly in the case of Fe/Gr/Pt. In contrast, the value of the tensor reaches similar values with and without Gr when Co is the ferromagnet.

The enhancement of the response in the presence of Gr reported in Fig. \ref{fig:Theory} offers an appealing scenario to explain the differences  observed experimentally in the spin-to-charge conversion taking place in Ir/Fe/Gr/Pt compared to Ir/Co/Gr/Pt. Due to numerical constraints, our calculations were limited to a small supercell containing only 4 ML of Co or Fe and 6 ML of Pt, which is much smaller than the real system that exceeds 10 nm of active materials. Therefore, our calculations do not fully account for the contribution of the ISHE from both Pt and Ir, but provide a strong indication of the major role played by the interfacial SOC induced on Gr. 
As discussed above, in the experiments the total spin-charge conversion arises from the additive contributions of the ISHE of Ir and Pt, as well as the IEE at the double Rashba interface, (Fe,Co)/Gr/Pt. Qualitatively, by using either Co or Fe as the spin injector, the interfacial IEE either counters or boosts the ISHE from Ir and Pt, resulting in the giant enhancement reported experimentally in Ir/Fe/Gr/Pt. Besides, the large anisotropy in the spin-charge conversion we experimentally demonstrate in Fig. \ref{fig:SP}b can be understood by the strong anisotropy of Gr band structure. Indeed, Fig. \ref{fig:Theory}b clearly shows that close to Fermi level, the spin-orbit coupled Gr states are concentrated around the K and K' points, and absent along the $\Gamma-M$ high symmetry line. As a result, the double Rashba interface is mostly active along $\Gamma-K$.

Our results open the way for new and robust systems where the use of epitaxial 2D materials sandwiched between 3D systems acquire a strong global SOC, and, in turn, a giant spin-charge interconversion. The high conductivity of graphene compared to other 2D material  together with the induced high spin conversion is also a promising feature for the realization of low powerapplicationss. Furthermore, the anisotropy of the interconversion could allow for tuning at will the efficiency, opening its use for logic applications and unconventional computing. 

\backmatter

\bmhead{Supplementary information}

\bmhead{Acknowledgments}

This work was funded by the French National Research Agency (ANR) through the ‘Toptronic’ ANR JCJC project, grant ANR-19-CE24-0016-01  and by  the project “Lorraine Université d’Excellence” reference ANR-15-IDEX-04-LUE. It was also partially supported by ANR CONTRABAS project, grant ANR-20-CE24-0023, by ANR through the France 2030 government grants EMCOM (ANR-22-PEEL-0009), PEPR SPIN ANR-22-EXSP 0007 and ANR-22-EXSP 0009. A.P. was supported by the ANR ORION project, grant ANR-20-CE30-0022-01, A. M. by the Excellence Initiative of Aix-Marseille Université-A*Midex, a French ”Investissements d’Avenir” program. Devices in the present study were patterned at Institut Jean Lamour's clean room facilities (MiNaLor). These facilities are partially funded by FEDER and the Grand Est region through the RANGE project. HRTEM characterization was performed at Institut Jean Lamour's microscope CC-3M facilities. IMDEA team acknowledges support from the Spanish AEI/MICINN Projects PCI2019-111867-2 (FLAG-ERA SOgraphMEM), PID2021-122980OB-C52 (ECLIPSE-ECoSOx), CNS2022-136143 (SPINCODE), and the “Severo Ochoa” Programme for Centres of Excellence in R\&D (CEX2020-001039-S). J.-C. and A.G. acknowledge support from the “(MAD2D-CM)-UAM” project funded by Comunidad de Madrid, by the Recovery, Transformation and Resilience Plan, and by NextGenerationEU from the European Union.

\section*{Declarations}


\begin{itemize}
\item Competing interests: The authors declare no competing interests.
\item Availability of data: Data are available upon reasonable request to the corresponding author.
\item Code availability: The authors
will agree to share the code upon reasonable request. 
\item Authors' contributions. JCRS, P.P., SPW, R.G. and A.A. proposed the experiments. P.P. and JCRS supervised the project. JCRS and SPW setup the spin pumping experiments and designed the masks. IA, R.G., A.G., and P.P. grew the samples, performed structural characterization and analyzed the results. J.G. performed the HRTEM characterization and its analysis. A.A. patterned the samples and performed most of the spin pumping measurements. JCRS analyzed the spin pumping measurements with inputs from A.A and SPW. SPW developed the expression to account the rf strength field in the spin pumping devices. A.P. and A.M. developed the theoretical model and ab-initio calculations. A.A., A.P., I.A., P.P., A.M., and JCRS wrote the manuscript with inputs from all the authors. 
\end{itemize}

\noindent


\begin{appendices}



\section{Methods}\label{sec:methods}

\bmhead{Sample growth}

The samples incorporating Gr (\textit{i.e.}, Gr/Pt or Gr/Al) and those without Gr (i.e., Pt or Al) were all fabricated in-situ on epitaxial Ir(111)/Fe(111) grown on sapphire crystals under controlled conditions, ensuring similar structural quality and clean interfaces. The methodology described in \cite{Ajejas2018,Ajejas2020} was followed. To begin, a 10 nm thick epitaxial Ir(111) layer was deposited on Al$_2$O$_3$(0001) single crystal substrates via DC-sputtering at 670 K with a partial Ar pressure of $8\cdot10^{-3}$ mbar and a low deposition rate (0.3\AA/s). Subsequently, for the Gr-based heterostructures, a monolayer of Gr was prepared by chemical vapor deposition through ethylene dissociation at 1025 K under a partial pressure of $5.5\cdot10^{-6}$ mbar. Then, Fe was deposited at room temperature using molecular beam epitaxy. The intercalation of Fe below Gr was facilitated by a thermal annealing process at 550 K. This procedure leads to the formation of a homogeneous Fe layer with high structural order and well-defined interfaces, similarly to the case of Co intercalation\cite{Ajejas2018,Ajejas2020}. For samples without Gr, a 7 nm-thick Fe layer was deposited via DC sputtering at RT directly onto the Ir(111) buffer. Finally, in all samples, a 5 nm capping layer of Pt or Al was DC sputtered at RT.

\bmhead{Device fabrication} 

The devices dedicated to spin pumping measurements were fabricated utilizing conventional UV lithography. The complete stack was patterned and then ion milled until reaching the substrate, controlled by an ion mass spectrometer employing a 4-wave IBE14L01-FA system. Subsequently, a second lithography step was done to grow a 200 nm insulating SiO$_2$ layer using RF sputtering via an Si target and Ar$^+$ and O$^{2-}$ plasma in a Kenositec KS400HR PVD. In a third step, contacts were then patterned and evaporated using an evaporator PLASSYS MEB400S. The dimensions of the active bar for spin pumping devices were 10 $\times$ 600 $\mu$m. The uniformity of device geometry, including SiO$_2$ thickness, coplanar waveguide dimensions and lateral dimensions of the milled samples, was maintained consistently across all the devices for a reliable comparison of SP-FMR voltages.

\bmhead{Structural characterization}

The systematic structural characterization of the samples by x-ray diffraction was performed using a commercial Rigaku SmartLab SE multipurpose x-ray diffractometer, equipped with a non-monochromatic Cu K$\alpha$ source ($\lambda$=1.54 \text{\normalfont\AA}).

Transmission electron microscopy (TEM) analysis was conducted using a JEM-ARM 200F Cold Field Emission Gun TEM/STEM (Scanning TEM) operating at 200 kV. The microscope is equipped with a GIF Quantum 965 ER and features a spherical aberration probe and image correctors, providing a point resolution of 0.12 nm in TEM mode and 0.078 nm in STEM mode. High-resolution TEM (HR-TEM) imaging was employed to investigate the atomic structure of the deposited layers. Furthermore, systematic energy dispersive spectroscopy (EDS) mapping analyses were systematically performed on the various samples.

\bmhead{Spin pumping measurements}
We performed experiments to measure SP-FMR using a probe station equipped with an in-plane DC magnetic field (\textit{H}) up to 0.6 T generated by an electromagnet. The schematic diagram of the SP-FMR devices used in these experiments is depicted in the inset of Fig. \ref{fig:SP}a. During the SP-FMR measurements, a fixed RF frequency current (\textit{f}) in the GHz range is injected into the coplanar waveguide in the device, generating an RF magnetic field (\textit{h}\textsubscript{RF}) on the sample. We sweep the DC external magnetic field and, when the system reaches the resonance condition, the magnetization of Fe precesses and leads to the generation of a spin current injected from Fe into the attached systems (Ir, Pt, Gr/Al, or Gr/Pt). Microscopically, the spin current injected into the adjacent layers comes from the spin precession and scattering of electrons at the interfaces, making this technique perfect for studying the spin current injection in our epitaxial Gr-based asymmetric stack. This spin current is subsequently converted into a charge current by ISHE in Ir or Pt, or by IEE at Rashba interfaces. In a open circuit, we measured the voltage (\textit{V}\textsubscript{SP}) by modulating the RF power injected into the coplanar waveguide using a lock-in synchronized to this modulation while sweeping the external \textit{H}. The power modulation employed a sine function with a depth of 100\% and a modulation frequency of 433 Hz. The resulting SP voltage peak exhibits a characteristic Lorentzian curve symmetric around the resonance field (\textit{H}\textsubscript{res}) when the system reaches the resonance condition, as illustrated in Fig. \ref{fig:SP}a. This curve was analyzed for different frequencies to determine the effective magnetization (\textit{M}\textsubscript{eff}) and the Gilbert damping ($\alpha$) of the ferromagnetic layer. More specifically, the center (\textit{H}\textsubscript{res}) and width ($\Delta H$) of the Lorentzian peak in \textit{V}\textsubscript{SP} were analyzed by fitting the voltage to a sum of a symmetric and an antisymmetric Lorentzian function \cite{Fache2020,Arango2022b}. Notably, in the SP measurements, the antisymmetric part of the signal was negligible, and only the symmetric part was considered in the fit. To calculate \textit{M}\textsubscript{eff}, the Kittel formula for an in-plane easy axis was employed (see e.g. \cite{Arango2022b} for more details). Additionally, the damping was determined by considering the linear dependence of $\Delta$H with frequency, where the Gilbert damping ($\alpha$) provided the slope, and the intercept, $\Delta H_0$, represented the frequency-independent inhomogeneous contribution \cite{Arango2022b}. Raw data of measurements on our different systems, as well as $M_{eff}$ and damping determinations are shown in Supplementary Information.

\bmhead{Isotropic damping and $I_c$ on Ir/Fe/Al and Ir/Fe/Pt}
In our Gr-free samples, we observe isotropic damping and charge current production (raw data and more details in Supplemental information). This is consistent with the isotropic result reported for the magnetic damping and spin-charge interconversion in epitaxial [001]MgO$||[001]$Fe$|$MgO, and [001]MgO$||[001]$Fe$|$[001]Pt \cite{Guillemard2018Charge-spinDirections}. Although spin-charge interconversion is anisotropic for [220]Pt along two no-equivalent in-plane directions \cite{Xiao_Fullerton}, it is isotropic for the [111]Pt in the thin-film growth direction as our case, Fig. \ref{fig:structure}a. Indeed, Gudin \textit{et al.} \cite{gudin2023isotropic} have shown that interconversion is isotropic along two no-equivalent in-planes directions for [111]Pt. In our Ir/Fe/Gr/Al sample, it is more complicated to detect whether there is an anisotropy due to the large linewidth and very low voltage or current signal.

\bmhead{Robustness and reproducibility} The spin pumping devices maintain their performance over time, tested up to two years after the initial measurements. This confirms on the one hand the robustness of the systems studied, in particular the samples with intercalated graphene, and on the other hand the reproducibility of our measurements. The signal measured even two years later shows the same amplitude and linewidth with a dispersion of less than 2\% for the very same devices. This also shows the reproducibility and reliability of the RF field strength value, $h_{rf}$, due to the microwave injected into the coplanar guide antenna in our experimental setup. Each sample has its integrated antenna, as everything is on-chip, but the geometry and the materials (Ti, Au, and SiO$_2$) used to establish such an antenna are the same for all the devices thus allowing a "direct" comparison of the charge current production in the different systems for the same RF conditions, \textit{i.e}, frequency and power injected in the antenna. Moreover, we have measured the DC spin pumping voltage in several devices within a 5x5 mm$^2$ block, and the results show similar signal for slabs patterning along $\Gamma-M$ (0, and 60 degrees) and $\Gamma-K$ (30 and 90 degrees) configurations as shown in the extended Fig 1. We can see that for Ir/Fe/Al there is practically no anisotropy, both, in amplitude and linewidth, for the 4 directions of high symmetries showed. In contrast, we confirmed that the results are totally different for ht emain system, Ir/Fe/Gr/Pt. Here, the sigbal are simialr for 0° and 60° which are equivalent directions in a hexagonal symmetry, but different for 30° and 90°. This implies that Gr/Pt imprints a strong SOC on the Fe layer (interface) that enforces the hexagonal symmetry of Gr. This is another confirmation of the result in Fig. 3d. Thus, it is plausible to consider that the SOC increases in Fe, in agreement with the theoretical results where it is seen that the first Fe layer shows an increase of SOC due to the presence of Gr and Pt. In addition, comparing the jump to zero field, off resonance, we will have another independent confirmation of our previous statement. This jump is observed in the Ir/Fe/Gr/Pt sample and not in the Ir/Fe/Al sample. The ANE and the SSE have the same symmetry. But such a jump is not observed in Ir/Fe/Al. Consequently, such off-resonance jump is dominated by the Spin Seebeck effect and its respective conversion of spin current into charge current and not by the ANE.    

\begin{figure}[H]
    \centering
    \includegraphics[width=0.85\textwidth]{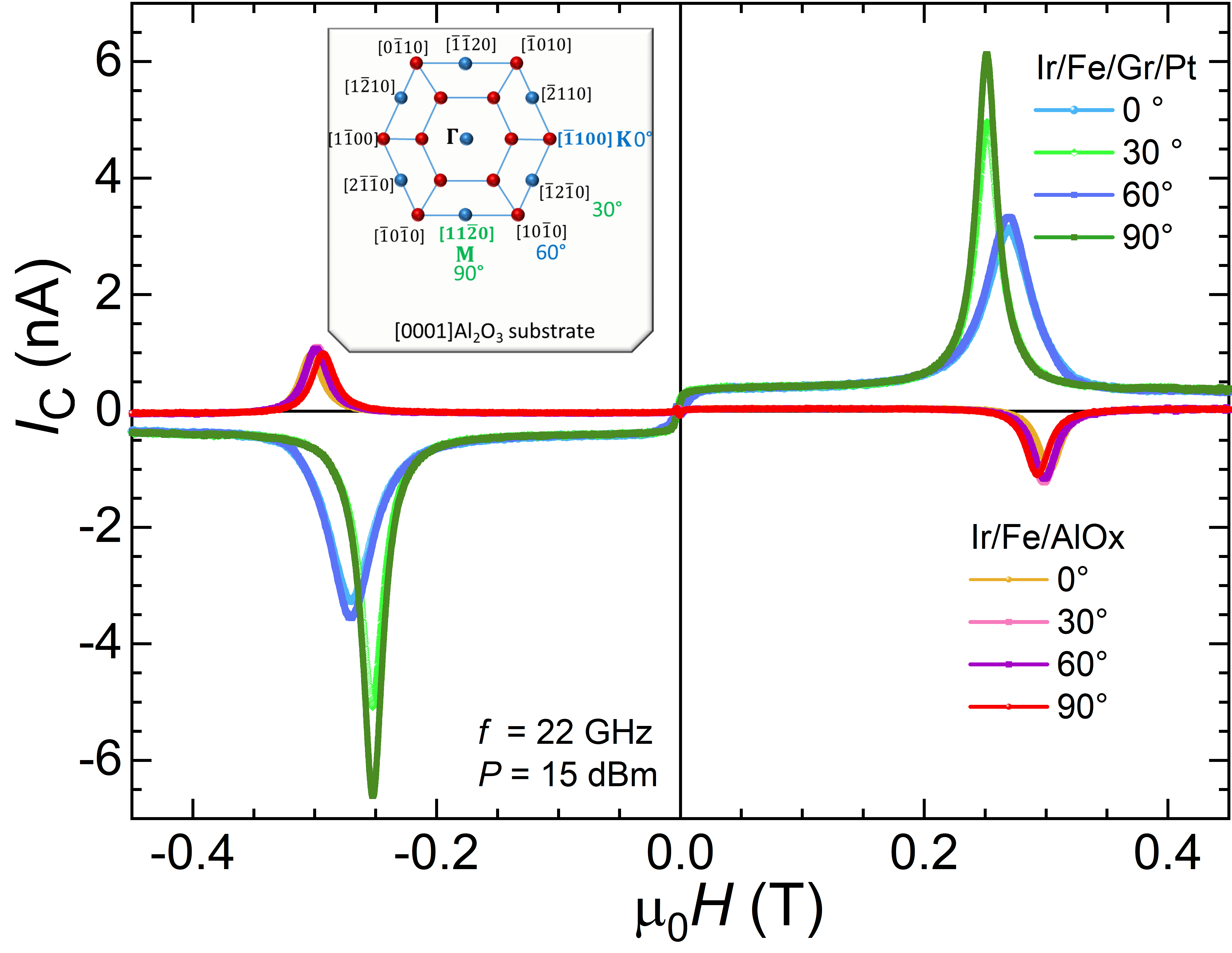}
      \caption{\textbf{Symmetry of spin pumping measurements.} (a) Raw data of charge current production by spin pumping at 22 GHz measured in slabs patterned every 30 degree following the directions of high symmetry according to the monocrystalline [0001]Al$_2$O$_3$ sapphire substrate as shown in the inset. We can see that Ir/Fe/Al does not exhibit anisotropy while Ir/Fe/Fe/Pt shows a repetition each 60 degrees in agreement with the hexagonal symmetry of Gr. Consequently, Gr/Pt imprints a SOC at the interface with Fe magnetic layer that dominates over the magnetocrystalline anisotropy of [110]Fe FCC. Comparing the jump to zero field off-resonance, we can conclude that it is dominated by the spin Seebeck effect (SSE). The anomalous Nernst effect (ANE) has the same symmetry as the SSE but such a jump is practically not present in the Ir/Fe/Al sample. This is another independent confirmation that Gr/Pt imprints a strong SOC on the Fe layer (at the interface).}
    \label{fig:Ic_0-90}
\end{figure}

\bmhead{Theoretical procedure}

{\it Realistic material simulations -} To compute the electronic and spin properties of the heterostructures, we consider trilayers having in total eleven layers, that is, four ferromagneticvlayers coming from bcc Iron or HCP Co, one Gr monolayer, and six Pt layers. Due to the lattice mismatch between the three materials, we have used a $2 \times 2$ Gr and Fe/Co supercell along with a $\sqrt{3}\times \sqrt{3}$ Pt supercell having a total of 42 atoms in the unit cell. We determined the band structure and spin textures by employing fully relativistic density functional theory \cite{siesta_method} within the On-site approximation to account for relativistic effects \cite{siesta_on-site_soc} introduced by fully relativistic pseudo-potentials along with the generalized gradient approximation (GGA) for the exchange-correlation functional \cite{pbe}. The calculations are converged for a 400 Ry plane-wave cutoff using the real-space grid with a $(13 \times 13 \times 1)$  $\vec{k}$-points sampling of the Brillouin zone. We used the conjugate gradient algorithm to minimize the atomic forces below 0.01 eV/{\AA}. 


\noindent The transport properties were calculated by a homemade procedure after extracting the Hamiltonian matrices from the self-consistent {\it{ab initio}} ground state. By using SISL as a post-processing, the velocity and spin projection matrices were obtained to perform the out-of-equilibrium simulatiosn.

\begin{table}
\caption{\label{table:mag_moment_graphene} Table showing the spin moment for each carbon atom in the Gr layer.}
 \begin{tabular}{ccc} 
Carbon atom & Fe & Co \\ 
 \hline
$C_1$  & -0.016 &  -0.094 \\
$C_2$  & -0.114 &  -0.095 \\ 
$C_3$  & -0.009 &  -0.093 \\
$C_4$  & -0.127 &  -0.092\\
$C_5$  & -0.018 &  0.023 \\
$C_6$ & -0.013 & 0.022 \\
$C_7$  & -0.127 &  0.018 \\
$C_8$  & -0.542 &  -0.294\\
 \end{tabular}
\end{table}

{\it Kubo-Bastin formula -}
 The current-driven field is computed within the linear response formalism considering the symmetrized decomposition of the Kubo-Bastin formula \cite{varga_2020} which has the form:

\begin{equation}
   h^{Ext}_{s_y}=-\frac{e\hbar}{8\pi}\int \partial_\epsilon f(\epsilon)d\epsilon \operatorname{Re}\left[\operatorname{Tr}\left\{\hat{v}_x(G^{\rm R-A})\hat{s}_y(G^{\rm R-A})\right\}\right]\label{ext}
\end{equation}
and 
\begin{equation}
    h^{Int}_{s_y}=-\frac{e\hbar}{4\pi}\int f(\epsilon)d\epsilon \operatorname{Re}\left[\operatorname{Tr}\left\{\hat{v}_x(G^{\rm R-A})\hat{s}_y(\partial_\epsilon G^{\rm R+A})\right\}\right],\label{int}
\end{equation}

\noindent where $\hat{v}_x=\partial_{k_x} \mathcal{H}$ is the velocity operator in the direction of the applied electric field $\hat{x}$, $\hat{s}_y$ is the spin matrix along $\hat{y}$ and $f(\epsilon)$ is the equilibrium Fermi distribution function. The function $G^{\rm R(A)}$ is the retarded (advanced) Green function and $G^{\rm R-A}=G^{\rm R}-G^{\rm A}$. The set of equations Eq. \ref{ext} and Eq. \ref{int} describe the extrinsic and intrinsic contributions to Spin-Orbit Torque. Analogously, the Rashba-Edelstein tensor is calculated using 

\begin{equation}
    \Lambda_{xy}=\frac{\sum_{n,\mathbf{k}}\braket{\hat{s}^y}_{n,\mathbf{k}}\braket{\hat{v}^{x}}_{n,\mathbf{k}}(\partial f_{n,\mathbf{k}}/\partial E_{n,\mathbf{k}})}{\sum_{n,\mathbf{k}}\braket{\hat{v}^{x}}_{n,\mathbf{k}}(\partial f_{n,\mathbf{k}}/\partial E_{n,\mathbf{k}})}.
\end{equation}





\end{appendices}

\bibliography{references,referencesADD}

\end{document}